\documentstyle[preprint,aps]{revtex}

\begin{document}

\draft

\title{ Probing quantum nanostructures
with near-field optical microscopy and (vice versa)}

 \author{Garnett W. Bryant}

\address{\it National Institute of Standards and Technology,
Gaithersburg, MD 20899, USA\\
e-mail: garnett.bryant@nist.gov}
\date{Current version dated \today }
\maketitle

 \begin{abstract} 
 \noindent A theory is presented to show how near-field optical 
microscopy can be used to probe quantum nanostructures. Calculations
are done for a quantum dot.
Results for different tip/dot configurations and sizes  
show that near-field excitation can enhance
light-hole transitions, excite selection-rule breaking transitions
with rates comparable to allowed transitions, and map electron-hole
pair wave functions. Conversely, dot response can be used to 
characterize tip near-fields. 
 \end{abstract}
 


Traditional far-field optical spectroscopy has been used extensively to
probe quantum nanostructures. Wavelength and polarization are varied to
access different transitions. However, spatial resolution, diffraction
limited to $\lambda/2$, is much larger than most 
nanostructures and only dipole allowed transitions can be excited.
Recently, near-field scanning optical microscopy (NSOM) has been
studied intensively to achieve spatial resolution much better than 
$\lambda/2$.  In NSOM, an optical fiber that has been tapered to 
a 10-100 nm wide tip and metal coated with an aperture at the tip is 
used as a localized light source or collector. NSOM has been exploited 
to investigate single molecules \cite{bc1,bian}, T-shaped quantum
wires \cite{grober}, and 10-nm wide self-assembled CdSe quantum dots 
\cite{flack}. In these cases the nanosystems were much smaller than the
NSOM probe field. NSOM spatial resolution was exploited only to
localize the excitation on a small collection of structures.
Development of theory for NSOM excitation of very small systems has 
begun \cite{hane,chang}.

Excitons confined at quantum well width fluctuations with a lateral 
size {\it comparable} to the probe field spread
have also been studied. NSOM was used 
to isolate single excitons in these systems, but spatial resolution 
was degraded by lateral exciton diffusion \cite{hess}.  
Isolated excitons were also studied by collecting photoemission that 
escaped through small holes in a metal layer on the surface of the 
quantum well sample \cite {zrenner,gammon}. In this case, the spatial
configuration of the aperture and confined exciton was fixed. No spatial
information about confined systems was obtained either in these cases.

In this letter, a simple theory is presented to show how NSOM can be used 
to probe nanostructures with sizes similar to the near-field spread.
As an example, a dot scanned by an NSOM tip is considered.
With NSOM, wavelength, polarization, {\it and} tip/sample separation can 
be used to control the transitions that are studied. The tip/sample 
separation can be varied to modify transition rates and access 
transitions that are forbidden in far-field spectroscopy. These effects 
depend on the probe-field spread being comparable to the nanostructure 
size. I will show that typical NSOM probes are sufficiently
localized to produce observable effects for confined systems.

A schematic of the experiment to be modeled is shown in Fig. 1. 
Simple but realistic models for the tip field, the dot, and
the dot confined states are used to reduce the complexity of 
the calculations and to focus on the essentials of this 
experiment rather than the details of a 
particular system. The tip field incident on the sample is described by
the Bethe-Bouwkamp model \cite{be,bouw}. In this approximation, the 
tip near-field is modeled by the near-field transmitted by
a circular aperture, radius $a$, in a perfectly conducting metal screen 
due to a plane-wave field incident normal to the screen. This model 
has been used successfully to understand NSOM measurements \cite {bc1,hsu}.
Implicit in this model is the assumption that local fields
due to the sample do not drastically alter the incident tip
field. The importance of local fields depends on the dielectric
contrast of the sample. The dot must be 
close to the top of the sample so that tip evansescent fields
can couple to the dot. Here, I assume that there is little 
dielectric contrast between the dot and its substrate.  
In that case, the substrate mainly refracts
the tip field. 

Optical transition rates are determined by Fermi's golden
rule using the spatially varying tip field to excite transitions.
The dot/probe coupling depends on the overlap of
the dot states with the probe field and on the atomic transition matrix
elements between confined electron and hole states. The dot is 
taken to be rectangular, with lateral widths $L_x$ and $L_y$ much greater 
than its thickness $L_z$. Infinite barriers are used to model the 
confinement. Using finite barriers instead  
would just increase the effective size of the dot states. 
Confined electron and hole envelop functions are found by use of 
single-band effective mass theory. The electron conduction 
states (e) are constructed from $s$ atomic states.
For the atomic state of the confined hole, I use the $J = 3/2$ atomic
states obtained with a 4-band Luttinger model for holes in a (001) 
quantum well with $k_x=k_y=0$. For heavy holes (hh), the atomic state 
is $\mp\sqrt{1/2}(|x\rangle\pm i|y\rangle)|\pm 1/2{\rangle}_s$; for
light holes (lh), $\sqrt{2/3}|z\rangle|\pm 1/2{\rangle}_s
+\sqrt{1/6}(|x\rangle\pm i|y\rangle)|\mp 1/2{\rangle}_s$. I assume that
the atomic state is unaffected by lateral confinement, which is 
reasonable for $L_x,L_y\gg L_z$.
The hh and lh response will be different, because the
lh contains a $|z\rangle$ component and can couple to electric
fields in the $z$-direction, $E_z$. $E_z$ is not present when probing
with a plane wave at normal incidence from the far-field, but is a 
significant component of the tip near-field. The 
optical excitations are electron-hole pair states formed from the 
confined single-particle states. In this paper I ignore 
pair interaction. These effects can be important in wide dots and will 
be included in a later study. 

To begin, it is
useful to remember how a dot responds to a linearly polarized 
plane wave incident normal to the dot from the far-field. 
Lateral quantum numbers are conserved ($n_{ex}=n_{hx}$ and 
$n_{ey}=n_{hy}$), because the incident field has no lateral
variation. The $z$ quantum number $n_z$ is conserved ($n_{ez} = n_{hx}$),
because $\lambda \gg L_z$. Also, the hh transition rate is 3 times 
the lh rate because there is no $E_z$ to couple to the lh.

To understand how NSOM can be used to probe nanostructures, one
must know the tip near-fields. In the Bethe-Bouwkamp model 
\cite {bc1,be,bouw}, a linearly polarized plane wave incident on 
an aperture drives charge around the aperture, producing an 
oscillating dipole moment in the direction of the incident 
polarization. The near field can be found in the
quasistatic limit and is the field of the instantaneous dipole. 
Near the aperture center and near the aperture plane, 
the near field is in the direction of 
the incident polarization (see Fig. 1). Away from the center there
is a weak field in the direction transverse to polarization and to $z$. 
Near the aperture edges, the dipole field is dominantly along
$z$. Thus the NSOM field has two large components; one 
along the incident polarization, which maintains its symmetry across 
the aperture but drops off rapidly at the edges, and $E_z$, which is 
large at the edges, small in the center, and changes sign going across the
aperture. Transitions that couple to $E_x$ and $E_y$
will be strongly excited when the tip is over the dot.
Transitions excited by $E_z$ will be stronger when the tip edge is
over the dot center. Transitions that normally are forbidden because
they do not conserve parity along the polarization can be excited
by $E_z$. Other symmetry-forbidden transitions become allowed when the tip 
is not centered over the dot. The coupling to $E_z$ and the forbidden 
transitions are more important for smaller apertures.
 
To show that these effects can be significant, I
present four examples. The field $E_i$ incident
on the aperture is $x$ polarized with wavelength $\lambda = 
822$ nm, corresponding to transitions near 1.5eV. The dot has 
$L_x = L_y = 100$ nm and $L_z = 10$ nm. Figures 2-5 present
rates for different transitions as a function of the tip/dot separation,  
defined in Fig. 1, as the tip is scanned at a height $z_t$ above 
the dot center line in the indicated direction. 

Figure 2 shows rates for transitions that conserve
quantum number with $n_x = n_z = 1$. When the tip
is over the dot center ($x = 0$), hh rates are 3
times the corresponding lh rates, as in
the far-field, because there is no coupling to $E_z$ when dot and tip
coincide. As the tip moves to the dot edge ($x = 50$ nm), hh rates
follow $E_x$ and decrease rapidly.
The lh rates respond to $E_z$ as well as $E_x$ and increase slowly 
until the tip edge passes beyond the dot center.
When the tip is outside the dot, lh transitions
are much stronger than hh transitions, reversing the usual ordering
of these rates. This reversal occurs where lh rates are
still significant and should be observable. When the tip scans
parallel to the incident polarization and probes 
states that vary monotonically (no nodes) away from the dot center in this
direction, the variations in the rates map the near field
and characterize tip size. In this case, the position of peak lh rate
indicates the tip radius.

Oscillations in pair-state wavefunctions can be probed
by scanning the tip along the oscillations. Rates 
for transitions with $n_x \geq 1$ are shown in Fig. 3 for the 
configuration used in Fig. 2, scanning along the polarization,
but with $z_t = 2.5$ nm
to increase coupling to the near field. 
Oscillations in the rates are due to 
variations in the near field along the polarization direction
and oscillations in the pair wave functions in this direction. To
separate these two effects, one can scan perpendicular to the polarization,
in the $y$ direction, where there are no oscillations in $E_x$ or $E_z$.
Fig. 4 shows a scan along $y$ for  
transitions with $n_x = 1$ but $n_y \geq 1$. As in Fig. 2,  
$z_t = 10$ nm, but now $a = 12.5$ nm so the spread 
of the near field is smaller than the dot and the tip acts 
more like a point source. Oscillations in the rates are now
due only to oscillations in the pair wave functions. These oscillations
are comparable to the magnitudes of the rates and should be observable. 
Oscillations are further enhanced by reducing $z_t$.

Scans done in different directions, for different polarizations, and
for different $z_t$ control how transitions are probed. 
Reducing $z_t$ increases the near field and the coupling to dot states 
when the tip is over the dot. Increasing $z_t$ spreads out the
tip field and enhances coupling to transitions which are important 
when the tip is outside the dot.

Transitions that do not conserve lateral
quantum number and would be weak in far-field spectroscopy can
be excited by the near field because the lateral spatial variation 
of the tip near field is comparable to the dot size. 
Transitions between confined states
with different lateral quantum numbers can be strongly excited
by a field with a lateral variation only if the lateral wavevector of the
field is so large that the field is evanescent along $z$
\cite {unpub}. Thus one must use near fields to excite these transitions.
Figure 5 shows rates for transitions
between the electron ground state and hole states with either
$n_{hx} > 1$ or $n_{hy} > 1$. The same configuration is
used as in Fig. 2.  Normally forbidden
transitions to $n_{hx} = 2$ hh and lh
states are strongly enhanced when the tip is moved from the dot
center to the peak in the hole wave function. Other scan
directions and configurations enhance other  
far-field forbidden transitions. These transitions
are weaker than number conserving transitions
(compare with Fig. 2). However, near the dot edge they are no more
than three times weaker and should be observable. Reducing $z_t$ increases
the rates for far-field forbidden transitions.  

In summary, a theory has been presented to show how near-field
optical microscopy can be used to probe nanostructures.
Scanning a near-field across a
nanostructure of similar size can selectively enhance
light hole transitions, selectively excite far-field forbidden transitions
with rates comparable to allowed transitions, and map electron-hole
pair wave functions. Conversely, the dot response can be used to 
map tip near-fields.


\begin{figure}
\caption{Schematic of a quantum dot scanned by an NSOM tip. Simplifications
in the theory are indicated. The aperture near field is
shown.}
\label{f1}
\end{figure}

\begin{figure}
\caption{Transition rates when an NSOM tip (radius $a = 50$ nm)
is scanned along $x$ at a height $z_t = 10$ nm above 
the dot center line. Quantum-number-conserving transitions with
$n_y \ge 1$ are shown.}
\label{f2}
\end{figure}

\begin{figure}
\caption{Transition rates when an NSOM tip is scanned along $x$ 
above the dot center line: $a = 50$ nm, $z_t = 2.5$ nm. 
Quantum-number-conserving transitions with $n_x \ge 1$ are shown.}
\label{f3}
\end{figure}

\begin{figure}
\caption{Transition rates when an NSOM tip is scanned along $y$
above the dot center line: $a = 12.5$ nm, $z_t = 10$ nm.
Quantum-number-conserving transitions with $n_y \ge 1$ are shown.}
\label{f4}
\end{figure}

\begin{figure}
\caption{Transition rates when an NSOM tip is scanned along $x$
above the dot center line: $a = 50$ nm, $z_t = 10$ nm.
Transitions between electron ground state and hole excited 
states are shown.}
\label{f5}
\end{figure}


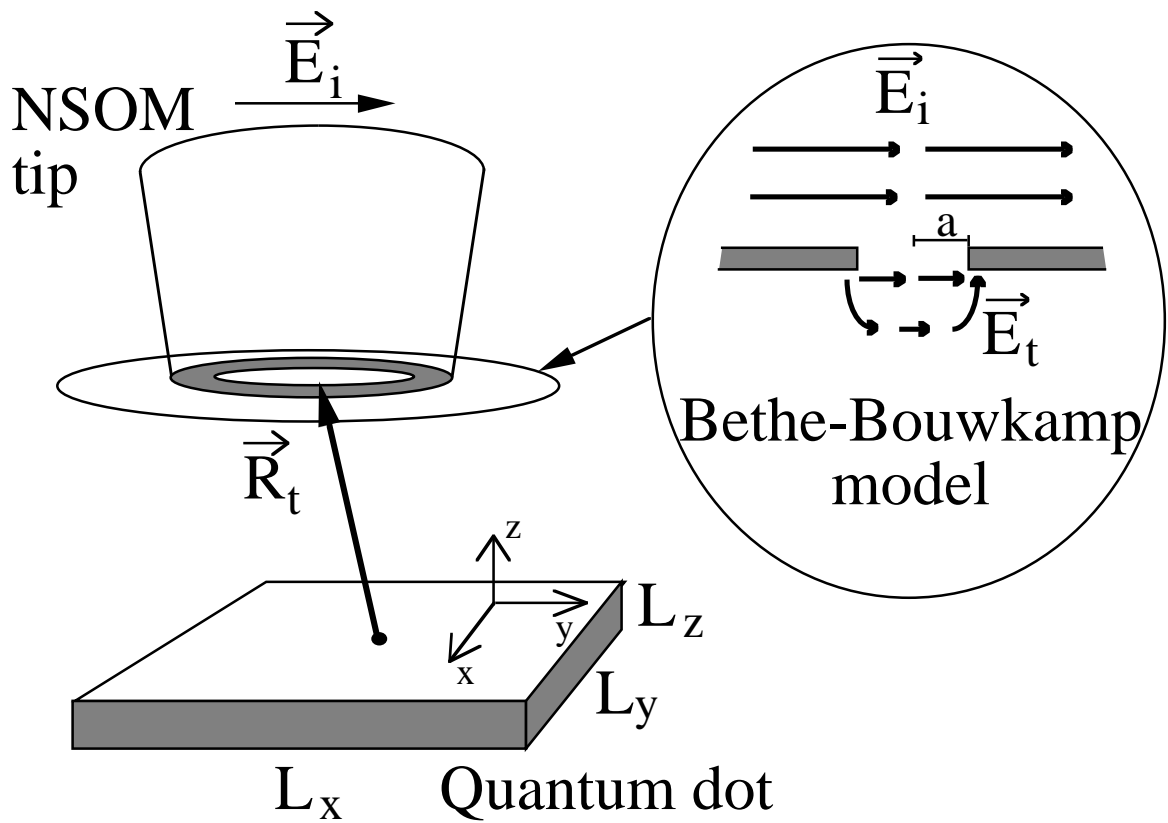
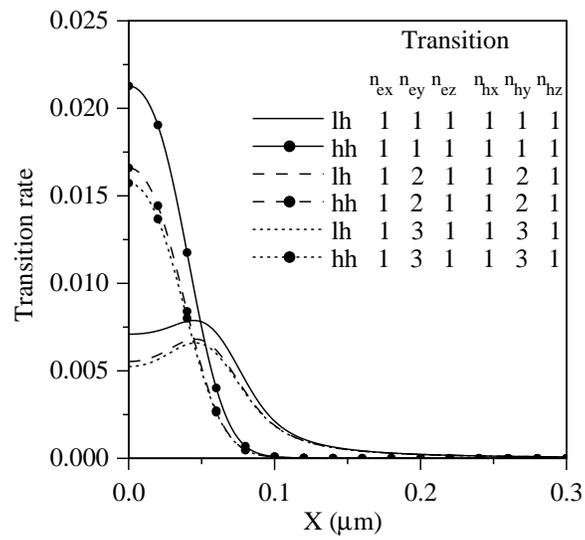
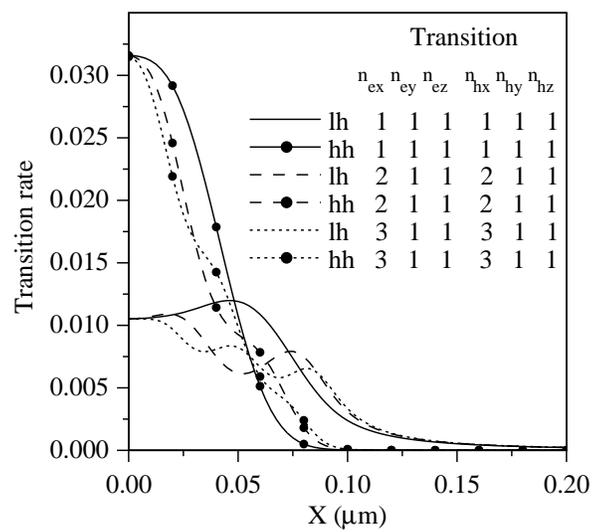
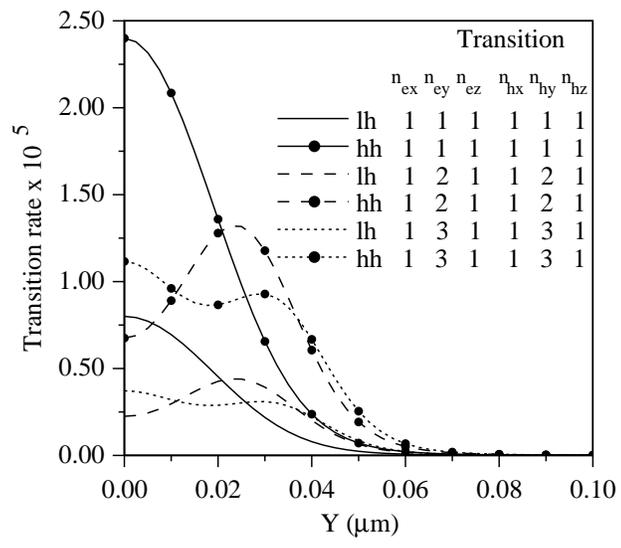
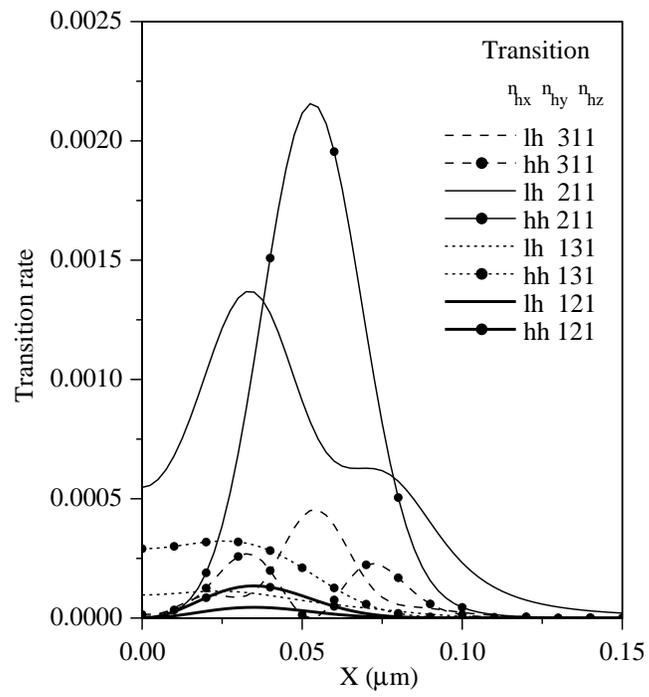
\end{document}